# MAPPING OF COVID-19 RISK FACTORS OF CITIES AND REGENCIES IN INDONESIA DURING THE INITIAL STAGES OF THE PANDEMIC

Setia Pramana, Department of Computational Statistics, Politeknik Statistika STIS, Indonesia, setia.pramana@stis.ac.id

Achmad Fauzi Bagus Firmansyah, BPS Statistics Indonesia, Indonesia, achmad.firmansyah@bps.go.id

Mieke Nurmalasari, Universitas Esa Unggul, Indonesia, mieke@esaunggul.ac.id

**Abstract:** The aims of this study are to identify risk factors and develop a composite risk factor of initial stage of COVID-19 pandemic in regency level in Indonesia. Three risk factors, i.e., exposure, transmission and susceptibility, are investigated. Multivariate regression, and Canonical correlation analysis are implemented to measure the association between the risk factors and the initial stage of reported COVID -19 cases. The result reveals strong correlation between the composite risk factor and the number of COVID-19 cases at the initial stage of pandemic. The influence of population density, percentage of people commuting, international exposures, and number of public places which prone to COVID-19 transmission are observed. Large regencies and cities, mostly in Java, have high risk score. The largest risk score owned by regencies that are part of the Jakarta Metropolitan Area.

**Keywords:** mapping risk, hazard, transmission risk, exposure risk, susceptible risk, COVID-19

## 1. INTRODUCTION

A group of cases of pneumonia with unknown cause occurred in Wuhan Hubei Province, China at the end of December 2019, and it was reported to China's National Health. In early January 2020, the virus was termed as the 2019 novel coronavirus (2019-nCoV), and it has been reported to WHO for the whole genome sequence.[1–3]

The 2019-nCoV cases not only occurred in Wuhan, in a short time nine cases were discovered in Thailand, Japan, Korea, the United States, Vietnam, and Singapore which is suspected of spreading by air travel. The rapid spread and the increasing number of cases were expected due to the celebration of the Chinese Lunar New Year. In addition, it is also predicted to around 15 million trips in Wuhan during the spring festival. These possibilities that make the virus quickly reached other places then become a pandemic.[4] The real-time updating for the distribution of the confirmed cases in the world can be seen in COVID-19 dashboard by Center for Systems Science and Engineering (CSSE) at Johns Hopkins University.[5]

The virus transmits from one individual to another via water droplets and physical contact. Coronaviruses are not airborne, these viruses spread rapidly through water droplets caused by sneezes and coughs. Inhalation of the water droplets transmits the virus into the human body. Physical contact is a common way of how the virus spreads.[6]

Touching a surface or an individual that is affected by COVID-19 could also transmit the virus through our body if mouth, eyes, and nose are touched after touching the Covid-19 affected surface.[7] Another study related to the transmission of coronavirus through human to human was also investigated by Ralph, R. et al.[8]

Population mobility plays an important role in the transmission rate of COVID-19. Demographics have changed, the mobility of people has increased, travel from foreigner and domestic countries has increased significantly, making it a challenge to mitigate the spread of the virus. Initial cases in many countries such as Thailand, Japan, Korea, and Italy are from travelers from China. Higher people mobility within and between regions increases the transmission of COVID-19. In high mobility regions, people are more likely to interact with other people thus transmitting the disease more quickly.[9] Furthermore, the movement of asymptomatic people are a potential source of corona virus infection with dynamic transmission.[10] Huang, R., was found that the infection spread of COVID-19 was spatially dependent, which largely spread neighboring areas from Hubei Province in Central China.[11]

Population density is also a key factor to the spread of COVID-19. People in areas with high population density have greater risks of being affected. Higher density means less room for physical distancing. Places with high density and mobility such as religious places, minimarkets, hotels, and schools are considered hazardous because the risk of getting COVID-19 is high. Population aging shows the risks for each age. Risks for older people tend to be higher than young people. This is likely associated by chronicle conditions at older ages. Comorbidity also could increase the chances of getting infected by COVID-19.[12,13]

In Indonesia, the government announced the first case of the corona virus in early March, 2020. On May 24, 2020, the Government of Indonesia declared 22,271 COVID-19 cases, 1,372 confirmed deaths and 5,402 cases recovered from 404 districts in 34 provinces.[14] Some areas have higher cases may due to their characteristics which can lead to higher risk of COVID-19. Knowledge understanding the risk factors can help to contain the spread of COVID-19 and lower the number of affected people.[15] Lummen & Yamada, (2014)[16], and Welle & Birkmann (2015)[17] indicate four main components (hazard, exposure, vulnerability, and capacity) for developing a disaster risk index.

The goals of this research are to identify the COVID-19 risk factor sets, construct a composite risk index, and map the cities and regencies in Indonesia based on the risk factor sets. The composite risk index is an indicator that reflects the risk degree of a region toward disaster including pandemic/epidemic. It is derived from three components, i.e., exposure, transmission, and vulnerability. The exposure factor indicates degree of exposure before pandemic occurred in a region. Transmission factor indicates degree of probability of a disease is being transmitted in a region. This index is related to the number of public facilities in a region. The vulnerability factor reflects degree of sensitivity in a region toward pandemic.

## 2. METHOD

### 2.1 Data

All risk factor variables/indicators are obtained from several survey carried out by BPS-Statistics Indonesia. The data is in regency's level, and there are 514 regencies all over Indonesia. Three risk factors were considered. First, the exposure factor which composes population density, proportion of commuters, and the number of foreign tourists. BPS Statistics Indonesia recorded number of tourists in several gates (airports, harbors, etc.) and border check points across Indonesia. For generating data in regency level, the buffering method of 25 km is used from the gates or check points. If the boundary of a regency is intersecting with the buffer, the number of foreign tourists in regency is the same as number of foreign tourists from gates of arrival, otherwise the number of foreign tourists is 0.

For both Ngurah-Rai and Soekarno-Hatta airports, we use buffer of 50 km instead of 25 km. The second factor is transmission and the indicator for this factor were worship places, number of mosques, churches, minimarkets, department stores or supermarkets or malls, banks, hotels, restaurants, and schools. The third is the susceptible factor. The indicators are age proportion above 50 years, proportion of people with morbidity, the proportion of the population who do not know to wash their hands properly, the proportion of the population who are commuting and sex ratio. Table 1 shows the variables used to calculate the composite risk factor of the COVID-19 pandemic.

| Factors | Variables | Sources |
|---|---|---|
| **Exposure** | Population Density | Statistics Indonesia (2019) |
| | Proportion of commuter | National Labor Force Survey (2018) |
| | Number of foreign tourist (January and February, 2020) | BPS Official Release Paper (February, 2020) |
| **Transmission** | Number of religious places | |
| | Number of minimarkets | |
| | Number of traditional markets | |
| | Number of supermarkets or malls | Indonesia Village Potency (2018) |
| | Number of banks | |
| | Number of hotels | |
| | Number of restaurants | |
| **Vulnerability** | Proportion of population who 50+ years old | Population Projection (March, 2020) |
| | Proportion of people have comorbidity | National Socio-economic Survey (2019) |
| | Proportion of population who do not have wash their hand well | Community Health Development Index (2018) |
| | Sex Ratio | Population Projection (2016) |
| | Percentage of household with house area < 8m$^2$ | National Socio-economic Survey (2019) |
| **Covid-19 Initial Cases** | Number of confirmed cases | Indonesia National Covid-19 Rapid Response Task Force |
| | Patients under surveillance (PDP) | |
| | People under monitoring (ODP) | |

**Table 1. The Risks Factors and the Initial Stage Covid-19 Hazard**

## 2.2. Risk Composite Index

In building a composite Index18 discussed several steps need to be performed. After theoretical framework and variable selection, normalization and, multivariate analysis and weighting and aggregation are the most crucial steps. Normalization is required prior to any data aggregation as the indicators in a data set often have different measurement units. It is done to render the indicators comparable and to account for extreme values and skewed data. There are several normalization methods exist.19,20 Welle and Birkmann (2015) transformed all indicators in dimensionless rank levels between 0 and 1 17. In this study as the data are in different unit and variability at the first stage all variables are normalized into values between 0 and 1 using the following equation:

$$z_i = \frac{(x_i - x_{min})}{(x_{max} - x_{min})} \quad (1)$$

The main challenge in building a composite index is to find the optimal weight for each indicator in the factors. Several weighting methods are available, principal components analysis (PCA) or factor analysis (FA) could be used to group individual indicators according to their degree of correlation [18]. In this study, the weights are calculated in two steps, first the weights of each indicator within factors using confirmatory factor analysis (CFA). The second is obtaining the weight of each factor in constructing the composite risk index. To get the weight that can optimal the correlation between the risk index and the number of COVID-19 cases, Canonical Correlation analysis is carried out.

Confirmatory factor analysis (CFA) is a method used to investigate and verify the shared variance of indicators that is believed to be attributable to a factor. [19] CFA is used to examine the hypotheses of the relationship between a set of variables with the latent variables. CFA model is constructed in advance specifies the number of latent factors and the patterns of loadings factors. CFA Model is defined:

$$x = \Lambda \xi + \delta, \qquad (2)$$

Where $x$ = (qx1) vector of indicator variable, $\Lambda$ = (q x n) matrix of factor loadings, $\xi$ = (n x 1) vector of latent construct (factors), and $\delta$ = (qx1) vector of errors of measurement. In this study, there are three factors with their corresponding indicators as defined in Table 1.

Before conducting the CFA, Bartlett's test of sphericity and Kaiser-Meyer-Olkin (KMO) Measure of Sampling Adequacy test are carried out to check whether the data is eligible for CFA analysis. The KMO indicates the proportion of variance of the variables/indicators that might be caused by underlying factors. High values (close to 1.0) generally indicate that a factor analysis may be useful with your data, whereas the value is less than 0.50 show that CFA is not very useful. Bartlett's test tests if correlation matrix is an identity matrix. It indicates if the variables are unrelated and therefore unsuitable for structure detection. Small values (less than 0.05) of the significance level indicate that CFA may be useful. To construct exposure, transmission, and susceptible factors, the weight of each indicators within the factors is obtained using formula:

$$w_{ik} = \frac{\lambda_{jk}}{\sum_{j=1}^{J} \lambda_{jk}} \cdot \frac{\sigma_{jk}^2}{\sum_{j=1}^{J} \sigma_{jk}^2} \qquad (3)$$

Where $w_k$ is the weight for *i-th* indicator of the *k-th* factor, $\lambda_{jk}$ and $\sigma_{jk}^2$ respectively are the CFA loading and the explained variance of the *j-th* indicator of the *k-th* factor.

After the risk factors have been obtained, the weight for each factor for constructing a composite risk index is calculated using Canonical Correlation Analysis (CCA). CCA is the analysis of association between two groups of variables. These variables are then used to construct Canonical Variates which are the weighted sum of the variables in the analysis.[20] Correlation is then being calculated between these variates and between the variables with the variates. For multiple x and y the canonical correlation analysis constructs two covariates (equation 4 and 5):

$$CV_{x_1} = a_1 x_1 + a_2 x_2 + a_3 x_3 + \cdots + a_n x_n \quad (4)$$

$$CV_{y_1} = a_1 y_1 + a_2 y_2 + a_3 y_3 + \cdots + a_m y_m \quad (5)$$

In this case, the X are all risk factors and Y are the number of confirmed cases, PDP, and ODP. The canonical weights $a_1...a_n$ and $b_1...b_n$ are chosen so that they maximize the correlation between the canonical variates $CV_{x_1}$ and $CV_{y_1}$.

The composite risk index is derived from three risk factors, i.e., Exposure, Transmission, and Vulnerability. The weight for each risk factor is calculated by the equation:

$$w_k = \frac{a_k}{\sum_{k=1}^{K} a_k} \qquad (6)$$

where, $a_k$ is the canonical weight of the *k-th* factor. Furthermore, early-stage hazard index is constructed from the number of confirmed cases, PDP, and ODP with their corresponding weight obtained from CCA.

The regencies are then grouped into five groups based on 1D K-means method of the risk index. The group is arranged so that it represents the ranks of the region. The higher rank the higher the risk. The risk rank of the regions is visualized in a map.

**2.3. Multivariate Multiple Regression**

The study also investigates the association between the numbers of cases at initial stages of COVID-19 with all indicators. The initial cases as the dependent variable consist of three variables, number of confirmed cases, ODP, and PDP, hence the most suitable method for this purpose is the multivariate linear regression. Multivariate multiple regression is a technique that estimates a single regression model with more than one dependent variables, i.e., the number of confirmed cases, PDP and ODP. The predictor variables are all risk factor variables. Procedures for statistical inference in the multivariate linear model, however, take account of the fact that there are several, generally correlated, responses. Multivariate multiple regression is the development of multiple regression in which the dependent variable consists of many dependent variables. Multivariate multiple regression is the development of multiple regression in which the dependent variable consists of many dependent variables. We consider modeling the relationship between m response variables and predictor variables. Every dependent variable is supposed to follow its own regression model in the same predictors.

$$Y_1 = \beta_{01} + \beta_{11}z_1 + \cdots + \beta_{r1}z_r + \varepsilon_1 \qquad (7)$$

$$Y_2 = \beta_{02} + \beta_{12}z_1 + \cdots + \beta_{r2}z_r + \varepsilon_2$$

$$Y_m = \beta_{0m} + \beta_{1m}z_1 + \cdots + \beta_{rm}z_r + \varepsilon_m,$$

where $\boldsymbol{\varepsilon} = [\varepsilon_1, \varepsilon_2, \ldots, \varepsilon_m]'$ is an error term which has $E(\boldsymbol{\varepsilon}) = 0$ and $Cov(\boldsymbol{\varepsilon}) = \sum$, the terms of error related with different responses might be correlated. In order to test the hypothesis of no overall effect of each predictor to all dependent variables, multivariate analysis of variance (MANOVA) is carried out. To perform the analysis, and visualization several python packages, such as **numpy**, **pandas**, **Mapclassify, statsmodels** [21], **seaborn** [22], and **geopandas** [23] are used.

**Statsmodels**, a python package utilized for multivariate analysis. The calculation of Canonical and Multivariate Analysis of Variances (MANOVA) had been conducted using this packages. **Geopandas**, an extension of **Pandas** modules that had been developed for spatial data manipulation and visualisation. This package had been developed using the **Pandas** as base packages and enhanced with Python spatial packages (i.e. **shapely**, **fiona**, and **pyproj**). **Seaborn**, a statistical data visualisation based on matplotlib. This package had been used for visualization and ease the exploration data analysis. In addition, **Mapclassify**, a python package for enhancing the classification process/ data binning for choropleth mapping. This package is a part of PySAL the Python Spatial Analysis Library.

## 3. RESULT

### 3.1. Composite Risk Index

The result of Bartlett test ($\chi^2$= 6387.04, p-value < 0.0001) and KMO (0.84) shows that CFA suitable to be used for capturing the structured of the data. The weights of each indicators on the corresponding factor obtained from CFA is shown in Table 2.

The result shows that all indicators seem to have similar weights. The weight of each factor in constructing a composite risk index resulted from CCA in shown in Table 1. The highest weight is for the transmission factor, followed by exposure, whereas vulnerability factor has relatively small weight. The weights for early-stage hazard for Positive cases is 0.4285, for PDP is 0.0451 and for ODP is 0.5265.

| Risk Factors | Weight (CCA) | Indicators | Weights (CFA) |
|---|---|---|---|
| Exposure | 0.3994 | Population density | 0.3328 |
| | | Proportion of commuters | 0.3356 |
| | | Number of foreign tourists | 0.3317 |
| Transmission | 0.5434 | Number of religious places | 0.1452 |
| | | Number of minimarkets | 0.1444 |
| | | Number of traditional markets | 0.1410 |
| | | Number of supermarkets or malls | 0.1437 |
| | | Number of banks | 0.14195 |
| | | Number of hotels | 0.14194 |
| | | Number of restaurants | 0.1417 |
| Vulnerability | 0.0572 | Proportion of population >50 years old | 0.2026 |
| | | Proportion of people have comorbidity | 0.1938 |
| | | Proportion of population who do not have wash their hand well | 0.1996 |
| | | Sex Ratio | 0.2130 |
| | | Percentage of household with house area < 8 m2 | 0.1909 |

**Table 2. Weight for Indicators and Risk factors**

### 3.2. Risk Factors Associated to Initial Stage of Reported Cases

Many capitals of provinces such as Jakarta and surrounding cities, and Surabaya have high exposure risk because of higher population density, probability of incoming foreign tourists, and risk of commuter worker. For the transmission risk factor, southern regions of Jawa Barat (West Java) have a high transmission risk, such as Bogor, Sukabumi, and Cianjur. This higher transmission risk occurs because the regions have many public facilities compared to the other regions. This transmission risk can indicate a probability of higher transmission of virus or other pandemic sources. Meanwhile, on the vulnerability risk factor, several regions have higher susceptible degree if pandemic occurs. Several regions located in western of Jawa Timur (East Java) and eastern of Nusa Tenggara Timur.

The map of the regencies based on the rank of composite risk index Figure 1. Note that the choropleth map is created using the cluster of the index values. Values in each bin have the same nearest center of 1D k-means cluster. Hence, the higher the rank (dark red colored), the higher the risk. It can be seen that the higher risk areas are mostly in Java and some parts of Sumatra. Several regions, mostly are province's capitals, have higher risk score such as Jakarta, Surabaya, Bandung, Makassar, Medan, and Denpasar. Moreover, their neighboring cities such as Bogor, Bekasi, Tangerang, Sidoarjo, Cianjur, and Garut have a moderate risk score.

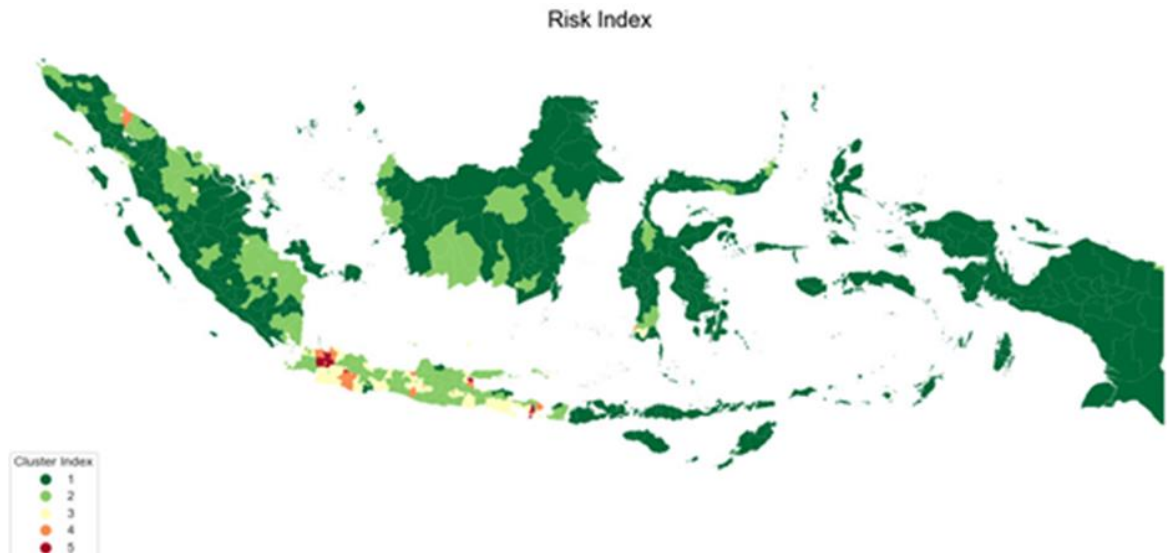

**Figure 1. Mapping of Composite Risk Factor**

Figure 2 present scatter plot of Covid-19 hazard against the risk factor all over Indonesia. It shows moderate correlation (0.695) between the composite risk factor and the initial reported cases (hazard). We observed that most of the initial confirmed cases are in Java. If we focus only for Java, the result show higher correlation 0.794. This is due the fact that at early stage of COVID-19 pandemic, regencies in Java are mostly affected with more cases compared to other regions.

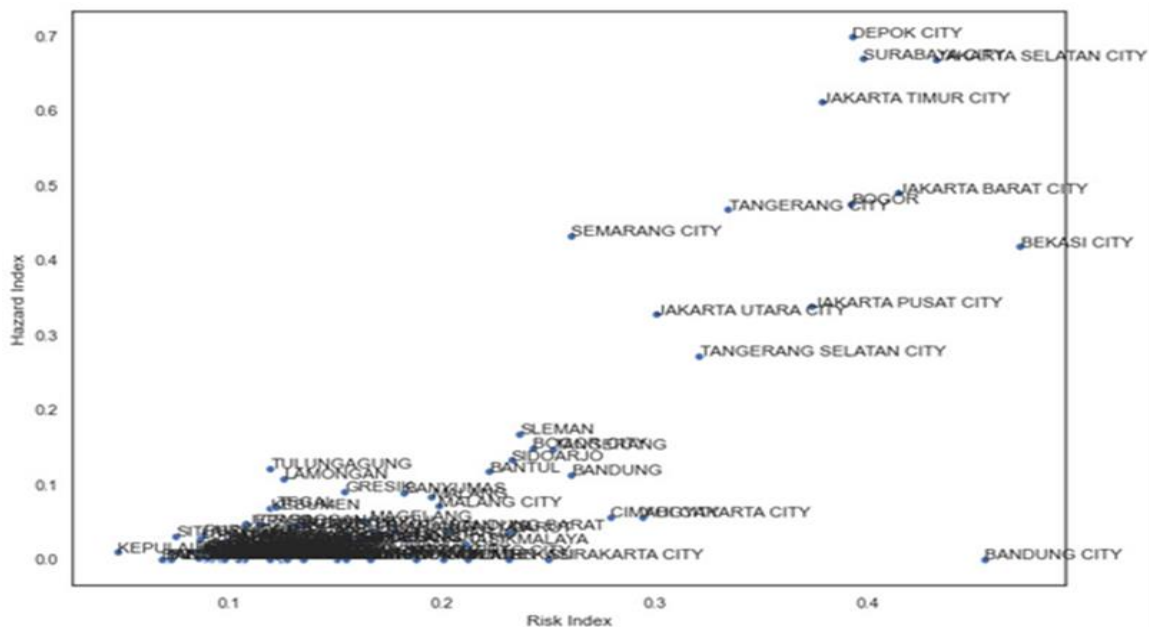

**Figure 2. Scatter plot of risk factors against initial stage of covid-19 hazard in Indonesia**

Figure 3 shows the composite risk factors of regencies in Java. The areas with higher exposure are the Jakarta Metropolitan (Jakarta, Bogor, Bekasi, Depok, Tangerang, and Tangerang Selatan) followed by other capitals such as Bandung, Semarang, Yogyakarta and Surabaya. These regions are business and governments area with high population density, large number of commuters, and international exposure. Whereas, the transmissions and vulnerability factors are more widespread, not only centered at the capitals or main cities. Higher risk regencies or cites are mostly located in the western part of Java.

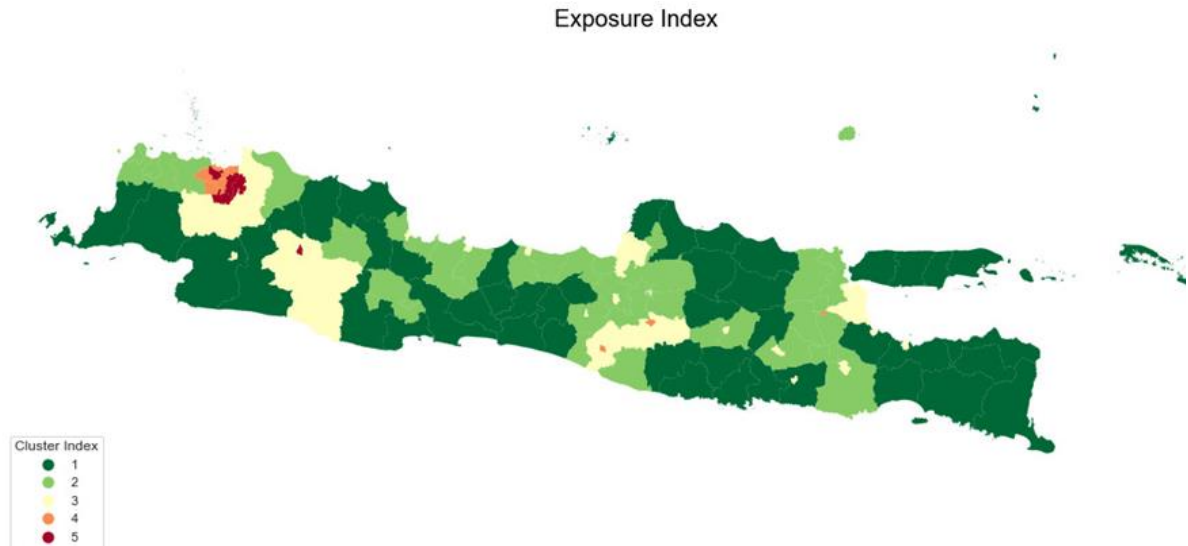

**Figure 3. Regency level risk factors mapping in Java**

### 3.3. Multivariate Regression Analysis

Some variables affect response variables differently. For example, confirmed: number of traditional markets and mini markets seem do not significantly related to the number of confirmed cases. Whereas, the number of ODP is only significantly related to the number of traditional markets, number of malls, and the proportion of population >50 years old. In addition, the number of PDP is not affected by the number of malls and mini markets. The results of MANOVA to test the overall effect of each indicator to all three dependent variables are presented in Table 3.

It shows that most of the variables of exposure risk factor are significantly associated with the number COVID-19 initial cases (confirmed cases, ODP and PDP). For the transmission factor, most of the indicators are also significant. Whereas, for the vulnerability factor only the proportion of households with house area < 8 $m^2$ is in statistically significant, and the proportion of population >50 years old is on border line.

## 4. DISCUSSION

The study shows that the highest composite risk score is Bekasi city, followed by Bandung, South Jakarta (Jakarta Selatan), Jakarta Barat (West Jakarta), and Surabaya. The top 10, except Bandung and Surabaya, are regencies included in the Jakarta Metropolitan Area. This shows substantial influence of international exposure on causing initial cases. This is why Jakarta Metropolitan area have the first outbreak and became the epicenter of the COVID-19 pandemic. Moreover, as the main region, with more densely populated and better ability to test and report cases, the number of reported cases grow sharply.

| Risk Factors | Variables | p-values |
|---|---|---|
| **Exposure** | Population density | 0.0033* |
| | Proportion of commuters | 0.0005* |
| | Number of foreign tourists | <0.0001* |
| **Transmission** | Number of religious places | 0.0004* |
| | Number of minimarkets | <0.0001* |
| | Number of traditional markets | 0.0146* |
| | Number of supermarkets or malls | 0.0004* |
| | Number of banks | <0.0001* |
| | Number of hotels | <0.0001* |
| | Number of restaurants | <0.0001* |
| **Vulnerability** | Proportion of population >50 years old | 0.0526** |
| | Proportion of people have comorbidity | 0.7442 |
| | Proportion of population who do not have wash their hand well | 0.1323 |
| | Sex Ratio | 0.8752 |
| | Percentage of household with house area < 8 m$^2$ | 0.0403* |

*Significant at α=5%, ** Significant at α=10%

**Table 3. Results of MANOVA testing for overall effects**

Notably, Bandung, Yogyakarta and Surabaya are cities that exhibit relatively high composite risk factor score. However, they have not seen the high levels of reported cases. This may show two possible explanations, the relatively effective efforts to contain the spread of the virus, or lagged reporting of cases. These regions are the main regions of five COVID-19 pandemic hotspot provinces in Indonesia, namely, Banten, DKI Jakarta, West Java, Central Java, and East Java reported by Eryando et. al. (2020).[24]

Furthermore, multivariate regression analysis shows that the number of reported cases (confirmed, ODP and PDP) for the onset of COVID-19 in Indonesia are mostly affected by the exposure factor (international exposure, population density, and people mobility), and transmission factor (e.g., number of banks, hotels, and restaurants).

Population density has been reported to be influence the propagation of COVID-19 in several countries such as Bangladesh[25] and Turkey[26]. Furthermore, several studies in different countries shows that people mobility is one of the key factors for the transmission[27–30]. In addition public spaces, such as offices[31] and other places[32] are reported to be key of transmission. Hence reduction of people mobility and number of visits to public places is substantial in reduction of the spread of COVID-19.

In this study, the number of reported cases may only reflect part of the visible problem, as it depends on the number of COVID-19 testing. The unidentified and unreported cases of COVID-19 may be more widespread as reported. It is either due to the lack of staff confidence in the initial handling of the reports or because of the overwhelmed data management structure at the local governments[33].

Furthermore, the composite index score was correlated with number of COVID-19 cases in different cities and regencies in Indonesia. The number of cases could vary across the regions depending on the level of restrictions imposed (mobility, lockdown etc.), COVID-19 surveillance system (testing, tracing, treatment facilities etc.) in the country.

Our study is focused on the early stage of pandemic where there are lack of pandemic management and policy. The result may not be directly applicable for currently situation where the pandemic management have been greatly improved and the policy is also much better. Our finding can be used as a basic of current or future risk factors development by updating the data and also adding more variables into the risk composite index. The inclusion of other more granular and real-time variables such as weekly mobility based on mobile phone, and GPS tracking (from volunteer with informed-consent) can greatly improve the quality and timeliness of the risk index.

## 5. CONCLUSION

The study has shown strong correlation between the exposure, transmission and vulnerability risk factors and the number of COVID-19 cases at the initial stage of pandemic in Indonesia. The influence of population density, percentage of people commuting, international exposures, and number of public places which prone to COVID-19 transmission are mostly affecting higher number of cases. Large regencies and cities, mostly in Java, have high risk score. The largest composite risk score are regencies included in the Jakarta Metropolitan Area. However, the number of reported cases and people under surveillance may be driven by the testing capacity and ability to trace the suspects. The subsequent phases are likely to include people mobility, the impact of large-scale restrictions in some areas, the local government actions, and also health capacity, such as number of COVID-19 referral hospitals, number of paramedics, number of ventilators, number of laboratories and its testing capabilities. The areas with low health capacity may suffer severe outbreaks. It is important to not only focus on regencies with high composite risk (e.g., densely populated cities), but also to the area where people tend to move from the epicenter due to economic reasons.

## REFERENCES AND CITATIONS